\documentclass[aps,prl,twocolumn,superscriptaddress]{revtex4-2}

\usepackage{graphicx,latexsym}
\usepackage{amsmath,amssymb}
\usepackage{amsthm}
\usepackage{bm}
\usepackage{braket}
\usepackage{longtable}

\usepackage{color}
\usepackage{comment}

\theoremstyle{remark}
\newtheorem{thm}{Theorem}

\usepackage[%
 setpagesize=false,%
 bookmarks=false,%
 bookmarksdepth=tocdepth,%
 bookmarksnumbered=true,%
 colorlinks=true,%
 linkcolor=blue,%
 citecolor=blue,%
 urlcolor=magenta%
]{hyperref}

\begin{document}


\title{Topological Characterization of Discrete-Time Classical Stochastic Processes:\\ Dual Role of Point-Gap Topology}



\author{Masaya Nakagawa}
\email{nakagawa@cat.phys.s.u-tokyo.ac.jp}
\affiliation{Department of Physics, The University of Tokyo, 7-3-1 Hongo, Bunkyo-ku, Tokyo 113-0033, Japan}
\author{Masahito Ueda}
\affiliation{Department of Physics, The University of Tokyo, 7-3-1 Hongo, Bunkyo-ku, Tokyo 113-0033, Japan}
\affiliation{Institute for Physics of Intelligence, The University of Tokyo, 7-3-1 Hongo, Bunkyo-ku, Tokyo 113-0033, Japan}
\affiliation{Fundamental Quantum Science Program (FQSP), TRIP Headquarters, RIKEN, Wako 351-0198, Japan}


\date{\today}

\begin{abstract}
We present topological characterization of classical stochastic processes described by discrete-time Markov chains on lattices. 
We point out that point-gap topology of stochastic matrices entails two distinct physical consequences that hinge on the choice of the reference point.
The point-gap topology around a generic reference point is related to the direction of transport, and nontrivial topology around the origin of the complex spectrum of a stochastic matrix implies non-Markovianity caused by, e.g., feedback control. On the basis of this characterization, we identify the topological origin of directed transport in a classic experiment of Maxwell’s demon [S.~Toyabe \textit{et al}., Nat. Phys. \textbf{6}, 988 (2010)] and find the topological nature of feedback control beyond thermodynamic interpretation. 
We demonstrate that a topologically enforced non-Markovian classical stochastic process can be simulated by a Markovian quantum master equation, indicating a topological form of quantum advantage.
\end{abstract}


\maketitle

Topology serves as a unifying framework in modern physics, where the wave nature of quantum systems gives rise to rich topological phases \cite{MoessnerMoore_book}. By leveraging techniques developed for topological quantum matter, similar topological phases have been realized in a variety of classical waves, including mechanical \cite{Ma19}, photonic \cite{Ozawa19}, and biological systems \cite{Shankar22}. Remarkably, systems of classical particles can exhibit topological phases in their stochastic dynamics \cite{Murugan17,Tang21,Sawada24_1,Sawada24_2,Agudo-Canalejo25} because probability distributions over space can be expanded in terms of eigenmodes, whose dynamics exhibit an effective wave-like character.

In contrast to static phases of matter, topological phases in nonequilibrium systems characterize global features of the dynamics through spectral topology of the time-evolution operator and its generator \cite{Gong18}. This idea has proven fruitful in periodically driven systems \cite{Oka09,Kitagawa10,Kitagawa10_2,Lindner11,Rudner13,RoyHarper17,Harper20,Rudner20}, non-Hermitian open systems \cite{Gong18,Kawabata19,Ashida20,Bergholtz21,Okuma23}, monitored quantum dynamics \cite{Oshima25,Xiao24}, and quantum feedback control \cite{Nakagawa25,Wen25}.
In this Letter, we develop a topological characterization of classical stochastic dynamics described by discrete-time Markov chains and find a dual role of their topology in transport and non-Markovianity.

Our approach is inspired by topological constraints on unitary quantum dynamics that prohibit topologically nontrivial time-evolution operators from being generated by local dynamics \cite{Kitagawa10,Gross12,Po16,Higashikawa19,Gong20,Liu25}. 
The key idea of our work is to impose topological constraints on stochastic dynamics to endow them with memory effects of past states.
Specifically, we show that nontrivial topology around the origin of the complex spectrum of a stochastic matrix hinders the corresponding local continuous-time Markovian dynamics. 
This implies that non-Markovianity in stochastic dynamics is caused by topological constraints.
In contrast to unitary operators whose eigenvalues are restricted to the unit circle in the complex plane, a stochastic matrix can also exhibit nontrivial topology around points other than the origin, leading to the unidirectional transport characterized by the nontrivial point-gap topology with respect to a generic reference point.

Our framework provides a topological characterization of the experimental realization of Maxwell’s demon in Ref.~\cite{Toyabe10} beyond the conventional information-thermodynamic interpretation  
and unifies distinct feedback protocols on the basis of their topological equivalence, which allows a considerable latitude in the way we control the system. 
We also show that appropriately designed feedback control can alter the topology of a stochastic matrix, thereby realizing operations that are impossible in Markovian dynamics without feedback control irrespective of its thermodynamic gain.
Furthermore, topologically nontrivial classical non-Markovian dynamics can be generated by a quantum Markovian master equation, demonstrating a decisive advantage of quantum systems in topological dynamics.

\textit{Topology of stochastic matrices}---We consider a discrete-time Markov process of a particle on a lattice. A state of the particle is specified by its position $j$ and internal degrees of freedom $a$ ($a=1,\cdots,D$). Let $\bm{p}(t)=(p_{j,a}(t))$ be the probability distribution of states of the particle at time $t\in\mathbb{Z}\tau$. The distribution at time $t+\tau$ is related to that at time $t$ as $\bm{p}(t+\tau)=T\bm{p}(t)$, where $T=(T_{i,a;j,b})$ is a stochastic matrix. 
The matrix element $T_{i,a;j,b}$ gives the transition probability from $(j,b)$ to $(i,a)$. 

Here we investigate the topology of stochastic matrices of discrete-time Markov chains in contrast to previous works on the topology of classical stochastic processes that focus on the generator of the master equation for continuous-time dynamics \cite{Murugan17,Tang21,Sawada24_1,Sawada24_2,Agudo-Canalejo25}. 
Since a stochastic matrix is non-Hermitian and has complex eigenvalues, we consider the point-gap topology of stochastic matrices \cite{Kawabata19}. If a stochastic matrix $T$ does not have an eigenvalue $\xi_{\mathrm{ref}}\in\mathbb{C}$ (i.e., has a point gap at $\xi_{\mathrm{ref}}$), a topological invariant of $T$ can be defined depending on the spatial dimension, symmetry, and the reference point $\xi_{\mathrm{ref}}$. 

We consider a homotopy between stochastic matrices. Two stochastic matrices $T_0$ and $T_1$ are homotopic if and only if there exists a continuous one-parameter family $T(s)$ of stochastic matrices that satisfies $T(0)=T_0$ and $T(1)=T_1$.
To make a physically meaningful classification of stochastic matrices, we need to impose additional constraints on homotopy. For instance, in the topological band theory, locality of Hamiltonians and the existence of an energy gap are assumed in most cases \cite{MoessnerMoore_book}. Here, we assume locality and invertibility of stochastic matrices $T(s)$ for any $s\in[0,1]$. The locality of a stochastic matrix $T=(T_{i,a;j,b})$ is defined as 
$T_{i,a;j,b}\leq C\exp\left(-|\bm{R}_i-\bm{R}_j|/\ell\right)$
for sufficiently large $|\bm{R}_i-\bm{R}_j|$ and some positive constants $C$ and $\ell$, where $\bm{R}_i$ is the coordinate of site $i$. The invertibility of $T$ implies the absence of zero eigenvalue, which is equivalent to the existence of a point gap at $\xi_{\mathrm{ref}}=0$. 
 
Suppose that two local and invertible stochastic matrices $T_0$ and $T_1$ are related to each other as
\begin{equation}
T_1=\mathcal{T}\exp\left[\int_0^1dt R(t)\right]T_0,
\label{eq_loc_gen_equiv}
\end{equation}
where $\mathcal{T}$ is the time-ordering operator,
for some rate matrix $R(t)$ that satisfies locality and generates a time-local master equation $\frac{d\bm{p}}{dt}=R(t)\bm{p}(t)$. Then,
\begin{equation}
T(s)=\mathcal{T}\exp\left[\int_0^sdt R(t)\right]T_0
\end{equation}
gives a homotopy between $T_0$ and $T_1$ since $\mathcal{T}\exp\left[\int_0^sdt R(t)\right]$ is local and invertible. 
Hence, if two stochastic matrices $T_0$ and $T_1$ are not homotopic, they cannot be written in the form of Eq.~\eqref{eq_loc_gen_equiv} for any local rate matrix $R(t)$ \footnote{A similar observation has been made for unitary quantum dynamics \cite{Gross12,Liu25}. However, the statement about non-Markovianity is unique to stochastic dynamics since unitary quantum dynamics is always memoryless. In addition, whereas the inverse statement, i.e., the existence of a generator that continuously connects two homotopic time-evolution operators, holds for the case of unitary quantum dynamics \cite{Liu25}, this does not necessarily hold in the present case because a rate matrix should satisfy additional constraints due to the conservation of probability.}. In particular, by setting $T_0$ to be the identity matrix, we obtain the following statement: if a local stochastic matrix $T$ is not homotopic to the identity matrix, it cannot be generated by any local Markovian master equation.
In other words, realization of such a topologically nontrivial stochastic matrix with a local continuous-time dynamics requires non-Markovianity that renders the stochastic matrix non-invertible at a certain time.

For translationally invariant stochastic processes, we can explicitly define a topological invariant that characterizes the topology of stochastic matrices. To make the discussion concrete, we consider a discrete-time Markov chain on a one-dimensional lattice with a stochastic matrix $T=(T_{i,a;j,b})\ (i,j\in\mathbb{Z})$ satisfying translational symmetry
$T_{i+1,a;j+1,b}=T_{i,a;j,b}$,
where we assume the periodic boundary condition (PBC) in the finite-size case.
The translational invariance allows us to perform the Fourier transformation of the stochastic matrix as
\begin{equation}
X_{a,b}(k):=\sum_lT_{j+l,a;j,b}e^{-ikl},
\end{equation}
where $-\pi\leq k\leq \pi$. We call the $D\times D$ matrix $X(k)=(X_{a,b}(k))$ the Bloch matrix \cite{Nakagawa25}, which is an analog of the Bloch Hamiltonian in the band theory. 

We utilize the locality of $T$ to define the winding number \cite{Gong18}
\begin{align}
w(T,\xi_{\mathrm{ref}}):=&\int_{-\pi}^\pi \frac{dk}{2\pi i}\partial_k \log\mathrm{det}[X(k)-\xi_{\mathrm{ref}}I]\in \mathbb{Z},
\label{eq_wind}
\end{align}
where $I$ is the $D\times D$ identity matrix.
In terms of the eigenvalues $\{\xi_n(k)\}$ of $X(k)$, we have
\begin{align}
w(T,\xi_{\mathrm{ref}})=&\sum_n\int_{-\pi}^\pi \frac{dk}{2\pi i}\partial_k\log (\xi_n(k)-\xi_{\mathrm{ref}}),
\label{eq_wind_xi}
\end{align}
which gives the total number of times the eigenvalues $\xi_1(k),\cdots,\xi_D(k)$ wind around $\xi_\mathrm{ref}$.
Since the winding number $w(T,0)$ is unchanged if $T$ is continuously deformed while keeping locality and invertibility, two stochastic matrices having different winding numbers $w(T,0)$ are not homotopic to each other and thus cannot be related to each other via Eq.~\eqref{eq_loc_gen_equiv} by using any local rate matrix. As a corollary, if a stochastic matrix has a nonzero winding number $w(T,0)$, it cannot be generated by any local Markovian master equation.

\begin{figure*}[t]
    \includegraphics[width=17.0cm]{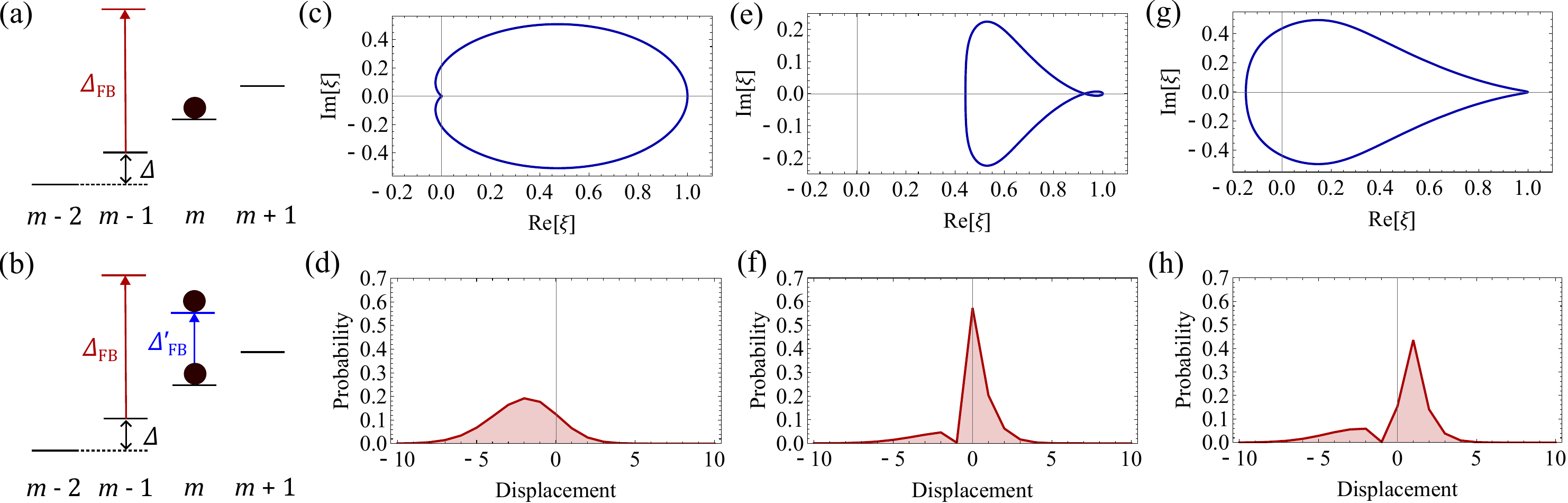}
    \caption{(a) Schematic illustration of the lattice model for the experiment in Ref.~\cite{Toyabe10}. A particle undergoes stochastic hopping on a one-dimensional tilted lattice with $\Delta$ being the potential difference between neighboring sites. A feedback potential $\Delta_{\mathrm{FB}}$ is applied on site $m-1$ if a particle is detected at site $m$. (b) Modified feedback protocol. An additional feedback potential $\Delta_{\mathrm{FB}}'$ is applied on site $m$. (c, e, g) Eigenspectrum of a stochastic matrix (c) without feedback control, (e) with feedback control, and (g) with modified feedback control. The parameters are set to $\beta\Delta=1,r=1,\tau=2$ [in (c), (e), and (g)], $\beta\Delta_{\mathrm{FB}}=5$ [in (e) and (g)], and $\beta\Delta_{\mathrm{FB}}'=2$ [in (g)]. (d, f, h) Probability distributions of the displacement of a particle from its initial position after time evolution during time $\tau$. Figures (d),(f), and (h) correspond to (c), (e), and (g), respectively. Note that the peak position of the distribution is negative for (d), zero for (f), and positive for (h).}
    \label{fig_Toyabe}
\end{figure*}

By way of illustration, we consider asymmetric random walk on a one-dimensional lattice, where the stochastic matrix is given by
\begin{align}
T_{i,j}=
\begin{cases}
    p_R & i= j+1;\\
    p_L & i= j-1;\\
    p_0 & i=j;\\
    0 & \mathrm{otherwise},
\end{cases}
\label{eq_stoc_discrete_walk}
\end{align}
with $p_R,p_L,p_0\geq 0$ and $p_R+p_L+p_0=1$. Since $\xi(k)=p_Re^{-ik}+p_Le^{ik}+p_0$, the winding number $w(T,0)$ of the stochastic matrix is well defined if $\xi(k)\neq 0$ for $-\pi\leq k\leq \pi$ and given by $w(T,0)=\mathrm{sgn}(p_L-p_R)\neq 0$ for $p_R+p_L>1/2$, showing that the discrete-time random walk with $p_R+p_L>1/2$ cannot be realized with any local Markovian master equation.

The problem about whether a given stochastic matrix can be realized with a continuous-time Markov process has been studied as the embedding problem for many decades \cite{Elfving37,Kingman62,Goodman70,Davies10,Baake20,Baake24}, but solved only for small system sizes. Our result shows that, by imposing locality on stochastic processes defined on lattices, topology offers a simple characterization of non-Markovianity. 
In the Supplemental Material \cite{supple}, we provide general properties of stochastic matrices concerning their topology characterized by the winding number, and also discuss the relationship between a topological invariant in zero-dimensional systems and existing results in the embedding problem \cite{Elfving37,Kingman62,Goodman70,Davies10,Baake20,Baake24}.

\textit{Topological nature of a Maxwell-demon experiment}---We apply the topological characterization of classical stochastic processes to the experimental realization of Maxwell's demon in Ref.~\cite{Toyabe10}.
We consider a particle on a tilted one-dimensional lattice [see Fig.~\ref{fig_Toyabe}(a) for a schematic illustration]. The particle has no internal degrees of freedom. 
A linear gradient potential is applied to the lattice and the potential difference between neighboring sites is denoted by $\Delta>0$. 
Without feedback control, the particle undergoes continuous-time asymmetric random walk governed by a master equation $\frac{d\bm{p}}{dt}=R\bm{p}$ with $r_{R}\ (r_L)$ being the transition rate to the right (left) neighboring site. We set $r_R=re^{-\beta\Delta/2}$ and $r_L=re^{+\beta\Delta/2}$,
where $r>0$ and $\beta>0$ is the inverse temperature of a heat bath, so that they satisfy the local detailed balance condition $r_R/r_L=e^{-\beta\Delta}$. 
We note that our topological characterization is applicable regardless of the local detailed balance.

Here, we impose the PBC to ensure the translational invariance of the model. Because of this boundary condition, the model does not satisfy the detailed balance condition $R_{i,j}\pi_{j}=R_{j,i}\pi_{i}$,
where $\bm{\pi}=(\pi_i)$ is the steady-state distribution defined by $R\bm{\pi}=0$. 
Although this violation of the detailed balance condition is an artifact due to the PBC, it does not affect transient behavior of the system if the particle is sufficiently away from the boundary. We note that the experiment in Ref.~\cite{Toyabe10} was performed in such a transient regime.

We perform discrete measurement-based feedback control of this system, which introduces a non-Markovian effect since it utilizes the information about measurement outcomes. We first measure the position of the particle, where we assume that the measurement is error-free. If the particle is found at site $m$, we raise the potential at site $m-1$ by $\Delta_{\mathrm{FB}}>0$ to prevent the particle from going to the left [see Fig.~\ref{fig_Toyabe}(a)]. The transition rates for the rate matrix $R_m$ under feedback control are modified accordingly in order to respect the local detailed balance condition. 
Let $\tau$ be the duration of feedback control, which corresponds to the interval between discrete times. Then the stochastic matrix of feedback control is given by $T=(T_{i,j})$ with
\begin{equation}
T_{i,j}=(e^{R_j\tau})_{i,j}.
\label{eq_stoch_matrix_fb}
\end{equation}

We first discuss the case without feedback control (i.e., $\Delta_{\mathrm{FB}}=0$). In this case, the rate matrix does not depend on the measurement outcome and therefore the stochastic matrix is given by $T=e^{R\tau}$. Figure \ref{fig_Toyabe}(c) shows the eigenspectrum $\xi(k)$ of the stochastic matrix. 
In this case, the winding number $w(T,0)$ vanishes, which is consistent with the fact that the stochastic matrix is generated by a master equation. 
Nevertheless, since the eigenspectrum forms a closed loop, it shows nonzero winding number $w(T,\xi_{\mathrm{ref}})=1$ for the reference point $\xi_{\mathrm{ref}}$ inside the loop.
The nonzero value of $w(T,\xi_{\mathrm{ref}})$ is related to directed transport in the model. Without feedback control, the particle is more likely to hop to the left due to the asymmetric transition rates $r_R<r_L$, which can explicitly be seen from the probability distribution $\{ T_{j+l,j}\}_l$ of displacement $l$ [see Fig.~\ref{fig_Toyabe}(d)]. If the eigenspectrum under the PBC has a positive (negative) winding number around some reference point, the eigenmodes of the stochastic matrix under the open boundary condition are localized near the left (right) edge of the system, which is known as the non-Hermitian skin effect in non-Hermitian topological systems \cite{Okuma20,Zhang20,Borgnia20}. In the present case, the non-Hermitian skin effect is a consequence of the directed transport due to asymmetric transition rates.

Next, we discuss the case with feedback control (i.e., $\Delta_{\mathrm{FB}}>0$). In this case, the stochastic matrix is given by Eq.~\eqref{eq_stoch_matrix_fb}. As shown in Fig.~\ref{fig_Toyabe}(e), the eigenspectrum of the stochastic matrix shows the winding number $w(T,\xi_{\mathrm{ref}})=-1$ for a reference point of, e.g., $\xi_{\mathrm{ref}}=0.5$. The sign reversal of the winding number indicates that the direction of transport is reversed due to feedback control, which can be verified from the probability distribution of displacement in Fig.~\ref{fig_Toyabe}(f). Thus, the effect of feedback control in this model can be topologically characterized by the winding number around some reference point.

However, as seen from Fig.~\ref{fig_Toyabe}(e), the winding number $w(T,0)$ vanishes even with feedback. This can be understood from the probability distribution of displacement. As shown in Fig.~\ref{fig_Toyabe}(f), the distribution takes its maximum value at $l=0$ and monotonically decreases as $l$ further increases, leading to the vanishing winding number (see the Supplemental Material for the proof \cite{supple}). 

A stochastic matrix with $w(T,0)\neq 0$ can be obtained if the feedback protocol is modified. 
The modified feedback protocol is schematically illustrated in Fig.~\ref{fig_Toyabe}(b). Suppose that a particle is detected at site $m$. In addition to the feedback potential $\Delta_{\mathrm{FB}}$ at site $m-1$, we introduce an additional feedback potential $\Delta_{\mathrm{FB}}^\prime$ at site $m$. The transition rates are modified accordingly so that the local detailed balance condition is satisfied. The eigenspectrum of the stochastic matrix, shown in Fig.~\ref{fig_Toyabe}(g), forms a loop enclosing the origin of the complex plane, giving the winding number $w(T,0)=-1$. The probability distribution of displacement $l$ in Fig.~\ref{fig_Toyabe}(h) takes its maximum at $l=1$, as expected from the fact that the particle is more likely to move to the right under the modified feedback protocol. Because of the nonzero winding number around the origin, the stochastic matrix of this feedback control cannot be generated by any local Markovian master equation that is unconditioned on measurement outcomes. 
Note that the topological distinction between Figs.~\ref{fig_Toyabe}(e) and (g) arises from the peak position and detailed structures of the probability distribution of displacement rather than thermodynamic gain discussed in Ref.~\cite{Toyabe10}.
Thus, the topological nature of feedback control beyond the conventional information-thermodynamic interpretation is demonstrated.

In the Supplemental Material \cite{supple}, we present additional numerical results to show the topological stability and equivalence of feedback protocols. 
Thus, the topology of stochastic matrices not only ensures topologically protected stability of the Maxwell-demon experiment but also leads to the topological equivalence between distinct feedback protocols in Ref.~\cite{Toyabe10}.

\textit{Classical versus quantum stochastic processes}---Having established the topological characterization of discrete-time Markov chains, we discuss the distinction between classical and quantum stochastic processes from a topological viewpoint. Here we consider a quantum generalization of the embedding problem proposed in Ref.~\cite{Korzekwa21}. A stochastic matrix $T$ is called \textit{quantum embeddable} if there exists a Liouvillian superoperator $\mathcal{L}(t)$ that satisfies 
\begin{equation}
T_{i,j}=\bra{i}\mathcal{E}(\ket{j}\bra{j})\ket{i},
\label{eq_q_embed}
\end{equation}
where $\{\ket{j}\}$ is an orthonormal basis set of the Hilbert space and $\mathcal{E}=\mathcal{T}\exp[\int_0^\tau dt\mathcal{L}(t)]$ is a quantum channel generated by a quantum master equation $\frac{d}{dt}\rho(t)=\mathcal{L}(t)(\rho(t))$. 
Physically, Eq.~\eqref{eq_q_embed} corresponds to the situation in which two-point projective measurements for the basis set $\{\ket{j}\}$ are performed before and after the quantum time evolution described by the quantum master equation.

\begin{figure}
    \includegraphics[width=8.0cm]{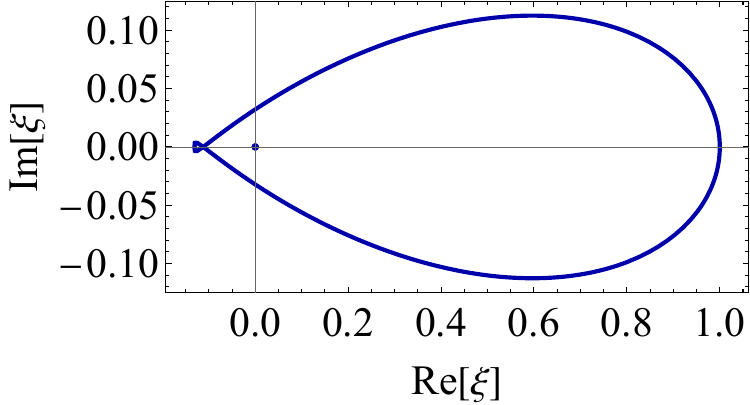}
    \caption{Eigenspectrum of the quantum channel in Eq.~\eqref{eq_q_embed_E}. The parameters are set to $J=0.8,\gamma_R=0.5,\gamma_L=0.25$, and $\tau=1$.}
    \label{fig_q_embed}
\end{figure}

Suppose that a stochastic matrix $T$ has a nonzero winding number $w(T,0)$ and thus cannot be generated by a local Markovian master equation. It is then natural to ask whether there exists a Markovian quantum dynamics with a local Liouvillian $\mathcal{L}$ that generates the stochastic matrix in the sense of Eq.~\eqref{eq_q_embed}. We answer this question in the affirmative. We consider a Liouvillian $\mathcal{L}$ in a one-dimensional lattice system where a quantum state at site $j$ is denoted by $\ket{j}$:
\begin{align}
    \mathcal{L}(\rho)=&-i[H,\rho]\notag\\
    &+\sum_{\alpha=R,L}\sum_{j}\gamma_\alpha\left(L_{\alpha,j}\rho L_{\alpha,j}^\dag-\frac{1}{2}\{ L_{\alpha,j}^\dag L_{\alpha,j},\rho\}\right).
\end{align}
Here,
$H=-J\sum_j(\ket{j}\bra{j+1}+\mathrm{H.c.})$
is the Hamiltonian of the system, and
$L_{R,j}=\ket{j+1}\bra{j}$ and $L_{L,j}=\ket{j-1}\bra{j}$
are Lindblad operators of stochastic hopping to the neighboring sites \cite{Haga21}. In Fig.~\ref{fig_q_embed}, we show the eigenspectrum of a quantum channel
\begin{equation}
    \tilde{\mathcal{E}}=\mathcal{E}_{\mathrm{proj}}\circ e^{\mathcal{L}\tau}\circ\mathcal{E}_{\mathrm{proj}},
\label{eq_q_embed_E}
\end{equation}
where $\mathcal{E}_{\mathrm{proj}}(\rho)=\sum_jP_j\rho P_j$ with $P_j=\ket{j}\bra{j}$ is the projective measurement channel. The eigenspectrum is composed of a closed loop and zero eigenvalues. The former is equivalent to the eigenspectrum of the stochastic matrix given in Eq.~\eqref{eq_q_embed} \cite{Nakagawa25}. The latter is due to the projective measurement described by $P_j$ since $P_j\ket{i}\bra{l}P_j=\delta_{i,j}\delta_{l,j}\ket{i}\bra{l}$. It is clear from Fig.~\ref{fig_q_embed} that the stochastic matrix has a nonzero winding number around the origin. Physically, as the zero eigenvalues are due to off-diagonal components of the density matrix in the $\{\ket{j}\}$ basis and absent in the classical case, this result is understood as a consequence of quantum coherence. 
This result indicates that quantum systems can simulate broader classes of topological stochastic processes than classical systems. 
Further details about the relationship between zero eigenvalue and Markovianity are given in the Supplemental Material \cite{supple}.

\textit{Conclusion}---
We have unveiled two distinct types of point-gap topology that are hidden in classical stochastic processes and characterized by two winding numbers around the origin and around a generic point in the complex plane. The former indicates the presence of non-Markovian effects and the latter reflects asymmetric transport. 
It merits further study to extend this framework by incorporating symmetry of classical stochastic processes \cite{Sa25}. For example, helical spin transport can be realized in a manner similar to that in Ref.~\cite{Nakagawa25} if reciprocal symmetry is imposed. From a broader perspective, it is of interest to understand how the topology of classical stochastic processes can be exploited to topologically constrain physical operations in stochastic thermodynamics \cite{Seifert}.

\begin{acknowledgments}
We are grateful to Akihiro Hokkyo, Atsushi Oyaizu, Makiko Sasada, Kazuki Sone, and Kazumasa A. Takeuchi for helpful discussions.
M.N.~was supported by KAKENHI Grant No. JP24K16989 from the Japan Society for the Promotion of Science (JSPS). 
M.U.~was supported by the RIKEN TRIP initiative, KAKENHI Grant No. JP22H01152 from JSPS, and the CREST program ``Quantum Frontiers" (Grant No. JPMJCR23I1) by the Japan Science and Technology Agency (JST).
This work was supported by JST as part of Adopting Sustainable Partnerships for Innovative Research Ecosystem (ASPIRE), Grant Number JPMJAP25A1.
\end{acknowledgments}

\bibliography{top_thermo_ref.bib}


\clearpage

\renewcommand{\thesection}{S\arabic{section}}
\renewcommand{\theequation}{S\arabic{equation}}
\setcounter{equation}{0}
\renewcommand{\thefigure}{S\arabic{figure}}
\setcounter{figure}{0}
\makeatletter
\c@secnumdepth = 2
\makeatother

\onecolumngrid
\begin{center}
\large{Supplemental Material for}\\
\textbf{``Topological Characterization of Discrete-Time Classical Stochastic Processes: Dual Role of Point-Gap Topology''}
\end{center}

\section{Properties of the winding number of stochastic matrices}

In this section, we show general properties of the winding number of stochastic matrices in one-dimensional stochastic processes.

\subsection{Symmetry and normalization}
We start with general properties of the Bloch matrix [see Eq. (3) in the main text for its definition]. 
Since a stochastic matrix is a real matrix, the Bloch matrix has the following symmetry:
\begin{equation}
X^*(-k)=X(k).
\label{eq_Bloch_Ksym}
\end{equation}
Because of this symmetry, eigenvalues of a stochastic matrix are real or appear as complex-conjugate pairs. In addition, since $\sum_{i,a}T_{i,a;j,b}=1$, the Bloch matrix satisfies
\begin{equation}
\sum_aX_{a,b}(k=0)=1.
\label{eq_Bloch_normalization}
\end{equation}

\subsection{Successive application of Markov processes\label{sec_successive}}
The Bloch matrix of a product of two stochastic matrices is given by the product of their Bloch matrices. Let $T_\alpha$ and $X_\alpha(k)$ $(\alpha=1,2)$ be stochastic matrices and their Bloch matrices. Then, the Bloch matrix of $T_1T_2$ is given by $X_1(k)X_2(k)$ since
\begin{align}
\sum_l(T_1T_2)_{j+l,a;j,b}e^{-ikl}=&\sum_l\sum_{r}\sum_{c}(T_1)_{j+l,a;r,c}(T_2)_{r,c;j,b}e^{-ik(j+l-r)}e^{-ik(r-j)}\notag\\
=&\sum_c(X_1(k))_{a,c}(X_2(k))_{c,b}\notag\\
=&(X_1(k)X_2(k))_{a,b}.
\label{eq_Bloch_product}
\end{align}
We use this property to obtain
\begin{equation}
w(T_1T_2,0)=w(T_1,0)+w(T_2,0).
\label{eq_w_T1T2}
\end{equation}
In particular, we have
\begin{equation}
w(T^n,0)=nw(T,0)
\end{equation}
for every nonnegative integer $n$. 
Thus, we can construct an infinite number of stochastic matrices with different values of the winding number from $T$ with $w(T,0)\neq 0$. 

In the following subsections, we will derive conditions for a stochastic matrix having vanishing winding number. By combining these conditions with Eq.~\eqref{eq_w_T1T2}, we obtain conditions for stochastic matrices with nonzero winding numbers. For example, the stochastic matrix $T^{(\mathrm{tr})}=(T_{i,a;j,b}^{(\mathrm{tr})})$ of one-site translation with $T_{i,a;j,b}^{(\mathrm{tr})}=\delta_{i,j+1}\delta_{a,b}$ has winding number $-D$, where $D$ is the number of internal degrees of freedom. Hence, if we multiply a stochastic matrix with zero winding number by $T^{(\mathrm{tr})}$, we can obtain a stochastic matrix with nonzero winding number $-D$.

\subsection{Parity formula}
The parity of the winding number is related to the sign of the eigenvalues at $k=0,\pi$:
\begin{equation}
    (-1)^{w(T,0)}=\prod_n\mathrm{sgn}[\xi_n(\pi)]\mathrm{sgn}[\xi_n(0)].
    \label{eq_sym_ind}
\end{equation}
To show this, we note that $\xi_n^*(-k)=\xi_n(k)$ holds due to the symmetry of the Bloch matrix in Eq.~\eqref{eq_Bloch_Ksym}. Because of this symmetry, $\xi_n(\pi)$ and $\xi_n(0)$ are real. 
Thus, we have
\begin{align}
    w(T,0)=&\sum_n\left[\int_0^\pi+\int_{-\pi}^0\right]\frac{dk}{2\pi i}\partial_k\log \xi_n(k)\notag\\
    =&\sum_n\int_0^\pi\frac{dk}{2\pi i}[\partial_k\log \xi_n(k)-\partial_k\log \xi_n^*(k)]\notag\\
    =&\sum_n\frac{1}{2\pi i}(\log\xi_n(\pi)-\log\xi_n^*(\pi)-\log\xi_n(0)+\log\xi_n^*(0)).
\end{align}
Since $\mathrm{Im}[\log \xi_n(k)]\equiv\mathrm{Arg}[\xi_n(k)]\ (\mathrm{mod}\ 2\pi)$ and $\mathrm{Arg}[\xi_n^*(k)]=-\mathrm{Arg}[\xi_n(k)]$, we obtain
\begin{align}
    (-1)^{w(T,0)}=&e^{i\pi w(T,0)}\notag\\
    =&\prod_n\exp\left[i(\mathrm{Arg}[\xi_n(\pi)]-\mathrm{Arg}[\xi_n(0)])\right]\notag\\
    =&\prod_n\mathrm{sgn}[\xi_n(\pi)]\mathrm{sgn}[\xi_n(0)].
\end{align}
A relation similar to Eq.~\eqref{eq_sym_ind} is known in the theory of symmetry indicators of band topology \cite{Po17,Po20}. In particular, if the Bloch matrix is one-dimensional (i.e., if the system has no internal degrees of freedom), we have $\xi(0)=1$ from Eq.~\eqref{eq_Bloch_normalization}. Thus, Eq.~\eqref{eq_sym_ind} reduces to $(-1)^{w(T,0)}=\mathrm{sgn}[\xi(\pi)]$. This result agrees with that of the asymmetric random walk in the main text.

\subsection{Enestr\"{o}m--Kakeya theorem\label{sec_EK}}
We consider the case with no internal degrees of freedom. In this case, the Bloch matrix is one-dimensional and eigenvalues of the stochastic matrix $T=(T_{i,j})$ are given by
\begin{equation}
\xi(k)=\sum_lT_{j+l,j}e^{-ikl}.
\label{eq_xi_singleband}
\end{equation}
In this subsection, we assume that the stochastic matrix is strictly finite-ranged, i.e., $T_{j+l,j}=0$ for sufficiently large $|l|$. 
Then, by using the argument principle, we can rewrite the winding number as
\begin{align}
w(T,0)=&-\oint_{|z|=1}\frac{dz}{2\pi i}\frac{\xi'(z)}{\xi(z)}\notag\\
=&N_{\mathrm{p}}-N_{0},
\label{eq_w_NpN0}
\end{align}
where
\begin{equation}
\xi(z):=\sum_lT_{j+l,j}z^l
\label{eq_xi_z}
\end{equation}
and $N_{\mathrm{p}}\ (N_0)$ is the number of poles (zeros) of $\xi(z)$ in $|z|<1$.

Suppose that 
$T_{j-l,j}=0$ for $l>l_1$, $T_{j+l,j}=0$ for $l>l_2$, $T_{j-l_1,j}\neq 0$, and $T_{j+l_2,j}\neq 0$ with $l_1$ and $l_2$ being nonnegative integers. If we set
\begin{equation}
\xi(z)=z^{-l_1}\tilde{\xi}(z),
\end{equation}
then $\tilde{\xi}(z)$ is a polynomial of $z$ with degree $l_1+l_2$, and the winding number is given by
\begin{equation}
w(T,0)=l_1-\tilde{N}_0,
\label{eq_w_l1N0}
\end{equation}
where $\tilde{N}_0$ is the number of zeros of $\tilde{\xi}(z)$ in $|z|<1$. Since the total number of zeros of $\tilde{\xi}(z)$ is $l_1+l_2$, we obtain bounds on the possible value of the winding number in terms of the range of the transition:
\begin{equation}
-l_2\leq w(T,0)\leq l_1.
\end{equation}

For the distribution of zeros of a polynomial, the following Enestr\"{o}m--Kakeya theorem holds (see, e.g., Ref.~\cite{Gardner14}):

\begin{thm}[Enestr\"{o}m--Kakeya]
If $p(z)=\sum_{l=0}^na_l z^l$ is a polynomial with real coefficients satisfying $a_0\geq a_1\geq\cdots\geq a_n\geq 0$, then all the zeros of $p$ lie in $|z|\geq 1$.
\end{thm}

Thus, if $\tilde{\xi}(z)$ satisfies the assumption of the Enestr\"{o}m--Kakeya theorem, we have $w(T,0)=l_1$. For example, if $T_{j+l,j}=0$ for $l<0$ and $T_{j,j}\geq T_{j+1,j}\geq\cdots\geq T_{j+l_2,j}$ (i.e., the transition is unidirectional and the transition probability monotonically decreases with increasing the travel distance), the winding number vanishes.

\subsection{Gershgorin circle theorem\label{sec_Gershgorin}}
Here we consider a general case with internal degrees of freedom. The Gershgorin circle theorem places a bound on the distribution of eigenvalues of a matrix \cite{MatrixAnalysis}:

\begin{thm}[Gershgorin]
Let $A=(a_{i,j})$ is a complex $n\times n$ matrix. For $j=1,\cdots,n$, we define a Gershgorin disc $D(a_{j,j},r_j)$ as a closed disc centered at $a_{j,j}$ with radius
\begin{equation}
r_j:=\sum_{i\neq j}|a_{i,j}|.
\end{equation}
Then, every eigenvalue of $A$ lies within at least one of the Gershgorin discs.
\end{thm}

We apply the Gershgorin circle theorem to the eigenvalue distribution of a stochastic matrix $T=(T_{i,a;j,b})$. Suppose that $T_{j,b;j,b}>1/2$ for all $j$ and $b$. Then, using $\sum_{i,a}T_{i,a;j,b}=1$ and $T_{i,a;j,b}\geq 0$, we have
\begin{equation}
\sum_{(i,a)\neq (j,b)}T_{i,a;j,b}<\frac{1}{2}<T_{j,b;j,b}
\end{equation}
for all $j$ and $b$.
Combining this inequality with the Gershgorin circle theorem, we find that all the eigenvalues of $T$ lie within the right half of the complex plane, which leads to $w(T,0)=0$. Thus, a nonzero winding number cannot be obtained if the particle stays in the same state with probability greater than $1/2$.

In the absence internal degrees of freedom, the above statement based on the Gershgorin circle theorem can also be made from Eq.~\eqref{eq_w_l1N0} and Rouch\'{e}'s theorem in complex analysis \cite{Ahlfors}. In fact, if $T_{j,j}>1/2$, we have
\begin{equation}
\Biggl|\sum_{l\neq 0}T_{j+l,j}z^{l+l_1}\Biggr|\leq\sum_{l\neq 0}T_{j+l,j}<\frac{1}{2}<T_{j,j}=|T_{j,j}z^{l_1}|
\end{equation}
for $|z|=1$. Thus, Rouch\'{e}'s theorem states that $\tilde{\xi}(z)$ and $T_{j,j}z^{l_1}$ have the same number of zeros in $|z|<1$, which leads to $w(T,0)=0$ from Eq.~\eqref{eq_w_l1N0}.

\subsection{Symmetric distribution}
Suppose that the transition probabilities have spatial inversion symmetry
\begin{equation}
\sum_{b,c}M_{a,b}T_{j-l,b;j,c}M_{d,c}=T_{j+l,a;j,d}\ (l\in\mathbb{Z}),
\end{equation}
where $M=(M_{a,b})$ is a real orthogonal matrix. In terms of the Bloch matrix, the inversion symmetry reads
\begin{equation}
MX(-k)M^\top=X(k).
\end{equation}
Then, we have
\begin{align}
w(T,0)=&\int_0^\pi\frac{dk}{2\pi i}\partial_k\log\det[X(k)]+\int_{-\pi}^0\frac{dk}{2\pi i}\partial_k\log\det[X(k)]\notag\\
=&\int_0^\pi\frac{dk}{2\pi i}(\partial_k\log\det[X(k)]-\partial_k\log\det[X(-k)])\notag\\
=&\int_0^\pi\frac{dk}{2\pi i}(\partial_k\log\det[X(k)]-\partial_k\log\det[MX(k)M^\top])\notag\\
=&0.
\end{align}
Thus, the inversion symmetry must be broken to obtain a stochastic matrix with a nonzero winding number. Combining this with the argument in Sec.~\ref{sec_successive}, we conclude that if the transition-probability distribution $\{ T_{j+l,a;j,b}\}$ is symmetric around $l=l_0$, the winding number is given by $-l_0D$.

\subsection{Detailed balance condition}
Let $\bm{\pi}=(\pi_{i,a})$ be a steady-state distribution of a discrete-time Markov chain with stochastic matrix $T=(T_{i,a;j,b})$: $T\bm{\pi}=\bm{\pi}$. The detailed balance condition is defined by
\begin{equation}
T_{i,a;j,b}\pi_{j,b}=T_{j,b;i,a}\pi_{i,a}\ (\forall i,j,a,b),
\label{eq_detailed_balance}
\end{equation} 
which means no net probability flow in the steady state. Equation \eqref{eq_detailed_balance} is rewritten as \cite{Sa25}
\begin{equation}
T\tilde{V}=\tilde{V}T^\top,
\end{equation}
where $\tilde{V}=\mathrm{diag}(\pi_{i,a})$. Here we assume that $\pi_{i,a}>0$ for all $i$ and $a$, which ensures the invertibility of $V$. We also assume the translation invariance $\pi_{i,a}=\pi_{a}$ of the steady state. Under these assumptions, the detailed balance condition can be expressed as symmetry of the Bloch matrix:
\begin{equation}
VX^\top(-k)V^{-1}=X(k),
\end{equation}
where $V=\mathrm{diag}(\pi_{a})$ is a $D\times D$ matrix.
Using this symmetry, we obtain
\begin{align}
w(T,0)=&\int_0^\pi\frac{dk}{2\pi i}\partial_k\log\det[X(k)]+\int_{-\pi}^0\frac{dk}{2\pi i}\partial_k\log\det[X(k)]\notag\\
=&\int_0^\pi\frac{dk}{2\pi i}(\partial_k\log\det[X(k)]-\partial_k\log\det[X(-k)])\notag\\
=&\int_0^\pi\frac{dk}{2\pi i}(\partial_k\log\det[X(k)]-\partial_k\log\det[VX^\top(k)V^{-1}])\notag\\
=&0.
\end{align}
Thus, a nonzero winding number requires the violation of the detailed balance condition; if a stochastic matrix has a nonzero winding number, its steady state must show nonvanishing probability flows.

\section{Topological invariant in the zero-dimensional case}
In the zero-dimensional case, a topological invariant of a stochastic matrix $T$ is given by (see Sec.~V of Ref.~\cite{Gong18})
\begin{equation}
s=\mathrm{sgn}(\det[T]).
\label{eq_detT}
\end{equation}
In fact, since eigenvalues of a stochastic matrix are real or appear as complex-conjugate pairs due to the symmetry in Eq.~\eqref{eq_Bloch_Ksym}, the quantity \eqref{eq_detT} corresponds to the even-odd parity of the number of negative eigenvalues of $T$, which cannot change unless $T$ has zero eigenvalue. As shown in the main text, a stochastic matrix with nontrivial topological index $s=-1$ cannot be realized with any Markovian master equation. This result reproduces a well-known fact in the Markov embedding problem that an embeddable stochastic matrix must have a positive determinant \cite{Elfving37,Kingman62,Goodman70,Davies10,Baake20,Baake24}.

\section{Vanishing winding number in single-site feedback control}
The stochastic matrix for the single-site feedback protocol in Fig.~1(a) in the main text has zero winding number around the origin, as seen from Fig.~1(e) in the main text. The reason can be understood from the probability distribution of displacement. As shown in Fig.~1(f) in the main text, the distribution takes its maximum at $l=0$ and monotonically decreases as $l$ increases for $l\geq 0$. Hence, if small probabilities for $l<0$ are neglected, the distribution satisfies the assumption of the Enestr\"{o}m--Kakeya theorem in Sec.~\ref{sec_EK}, and therefore the winding number vanishes. The vanishing winding number also follows from the Gershgorin circle theorem in Sec.~\ref{sec_Gershgorin} since $T_{j,j}>1/2$ as seen from Fig.~1(f).

\section{Stability of topological feedback control against imperfection}
Since the value of a topological invariant cannot be changed by small continuous deformation, 
feedback control with a stochastic matrix having nonzero winding number is expected to be stable against noise and imperfection during the dynamics. To corroborate this expectation, we study the effect of time delay in feedback control, which also exists in the experiment in Ref.~\cite{Toyabe10}. We consider the model on a tilted one-dimensional lattice considered in the main text. Let $\tau$ be the duration of time evolution and $\tau_{\mathrm{d}}$ be the delay time of feedback control. Then, the stochastic matrix of feedback control is given by
\begin{equation}
T_{i,j}=(e^{R_j(\tau-\tau_{\mathrm{d}})}e^{R\tau_{\mathrm{d}}})_{i,j},
\label{eq_T_delay}
\end{equation}
where $R$ is the rate matrix for the dynamics before the feedback control starts.

\begin{figure*}[t]
    \includegraphics[width=17.0cm]{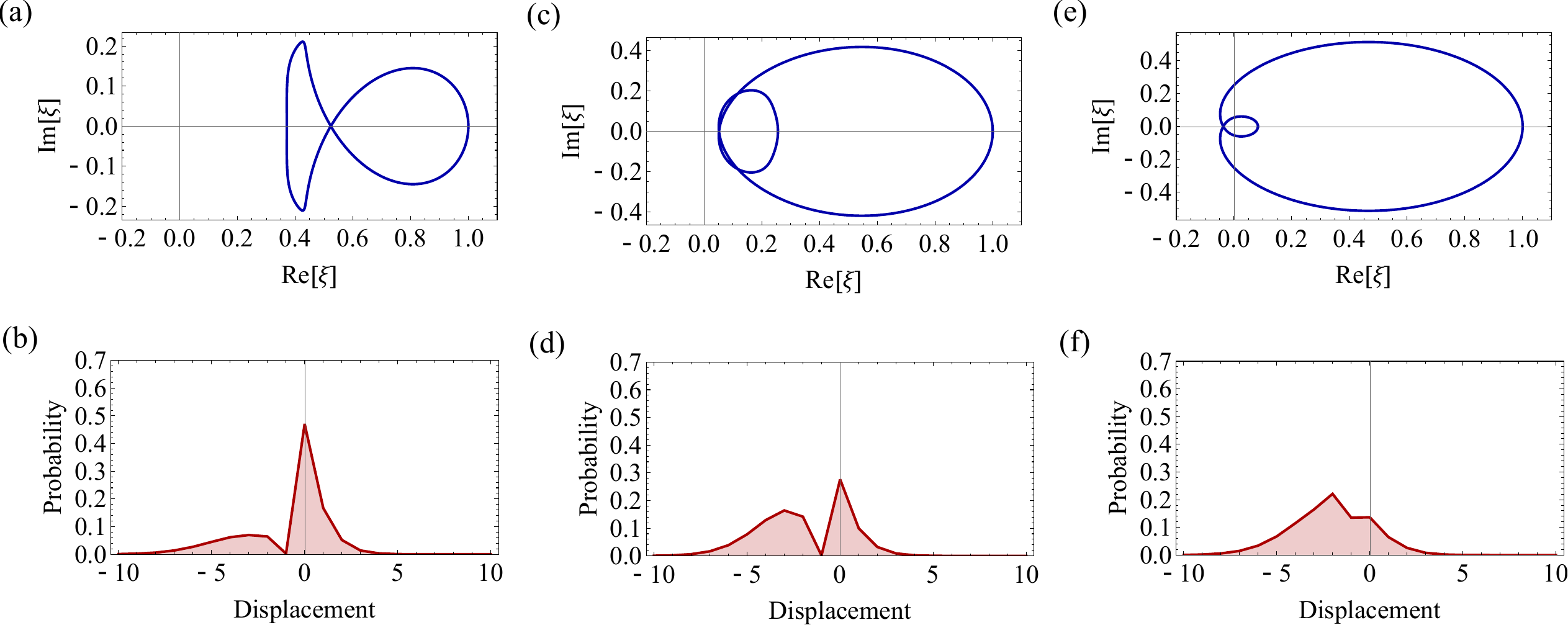}
    \caption{(a, c, e) Eigenspectra of a stochastic matrix under feedback control with time delay. The feedback protocol is the same as in Fig.~1(a) in the main text. The parameters are set to $\beta\Delta=1,r=1,\tau=2$, and $\beta\Delta_{\mathrm{FB}}=5$. The delay time is (a) $\tau_{\mathrm{d}}=0.2$, (c) $\tau_{\mathrm{d}}=1$, and (f) $\tau_{\mathrm{d}}=1.99$. (b, d, f) Probability distributions of displacement of a particle from its initial position after the time evolution during time $\tau$. Figures (b),(d), and (f) correspond to (a), (c), and (e), respectively.}
    \label{fig_Toyabe_delay}
\end{figure*}

In Figs.~\ref{fig_Toyabe_delay}(a), (c), and (e), we show the eigenspectrum of the stochastic matrix given in Eq.~\eqref{eq_T_delay} for several values of $\tau_{\mathrm{d}}$. We also display the corresponding probability distribution of displacement after the time evolution in Fig.~\ref{fig_Toyabe_delay}(b), (d), and (f). If the delay time is sufficiently short [e.g., in Fig.~\ref{fig_Toyabe_delay}(a)], the winding number $w(T,\xi_{\mathrm{ref}})$ around a certain reference point $\xi_{\mathrm{ref}}$ still has a negative value of $-1$ (for e.g., $\xi_{\mathrm{ref}}=0.4$) as in the case without time delay, indicating the topological stability of the feedback control. If the delay time is increased, a topological transition occurs and the winding number takes nonnegative values for any reference point [see Fig.~\ref{fig_Toyabe_delay}(c)]. Interestingly, if the delay time is further increased, the stochastic matrix has a positive winding number around the origin of the complex plane [see Fig.~\ref{fig_Toyabe_delay}(e)]. The nonzero value of the winding number around the origin implies that this stochastic matrix cannot be implemented with the Markovian dynamics and that the feedback control plays a decisive role even if the duration $\tau-\tau_{\mathrm{d}}$ of feedback control is small.

\begin{figure*}[t]
    \includegraphics[width=18.0cm]{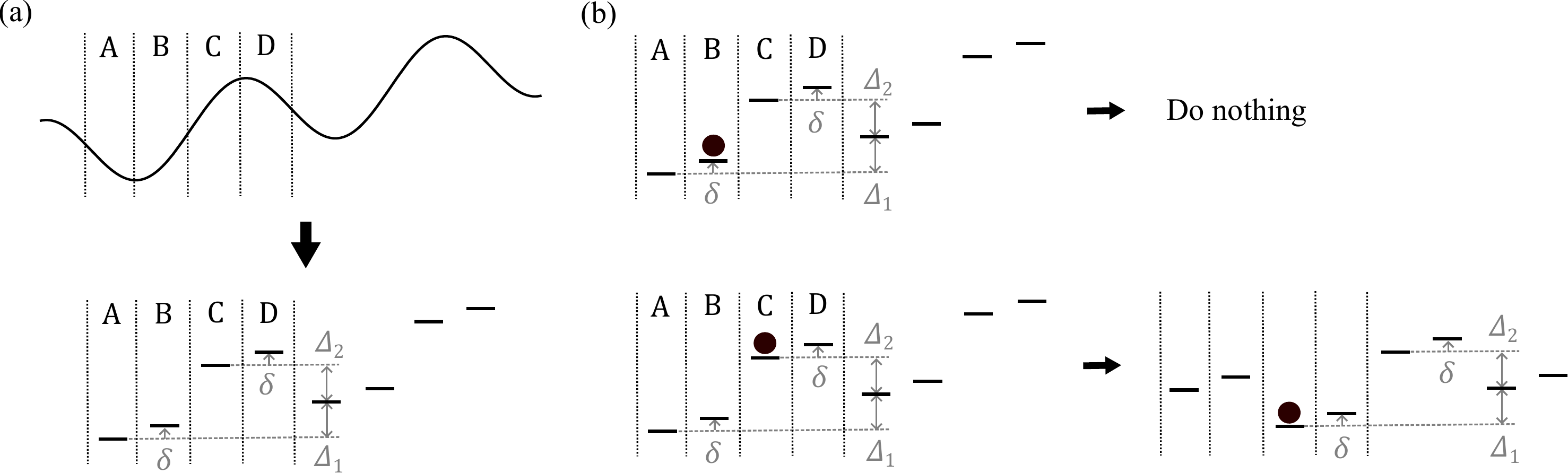}
    \caption{(a) Schematic illustration of the lattice model for the actual feedback protocol in the experiment in Ref.~\cite{Toyabe10}. The four regions of the periodic potential (upper panel) are represented as four sublattices in the lattice model (lower panel). (b) Protocol of feedback control. If a particle is found at sublattice C, the lattice potential is switched.}
    \label{fig_Toyabe_sublattice}
\end{figure*}

\begin{figure*}[t]
    \includegraphics[width=18.0cm]{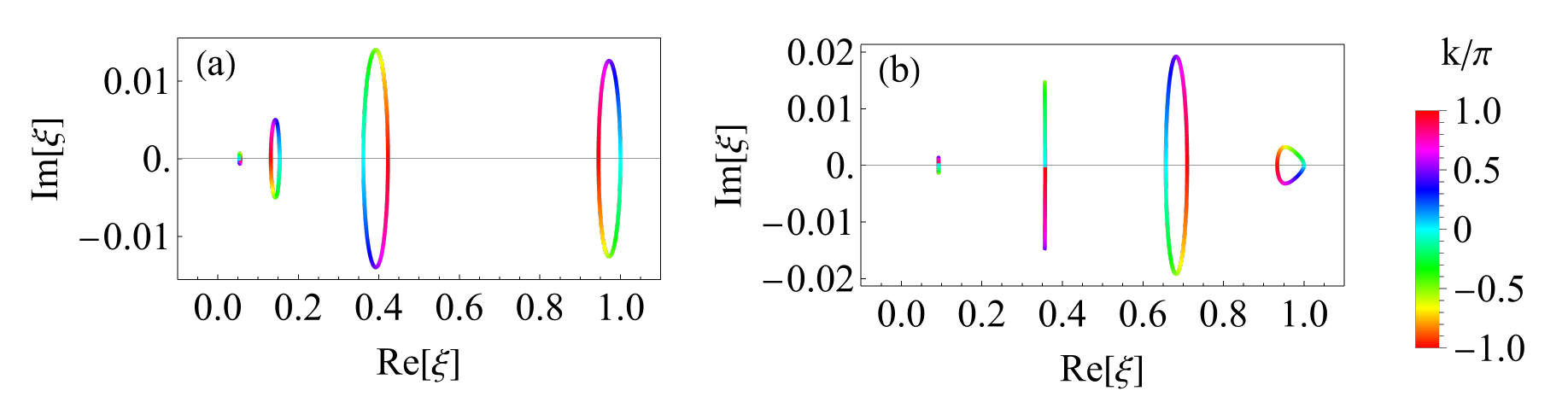}
    \caption{Eigenspectrum of a stochastic matrix for the four-sublattice model (a) without feedback control and (b) with feedback control. The parameters are set to $\beta\Delta_1=1,\beta\Delta_2=2,\beta\delta=0.25,r=1$, and $\tau=0.5$. The periodic boundary conditions are imposed, and the wavenumber of each eigenmode is shown by color according to the color bar on the right.}
    \label{fig_Toyabe_sublat_spec}
\end{figure*}

\section{Topological equivalence between different feedback protocols}
In the experiment in Ref.~\cite{Toyabe10}, the feedback control is performed by \textit{globally} switching a periodic potential when a particle is detected at a certain region. Although this protocol was schematically illustrated in Ref.~\cite{Toyabe10} where a wall potential is placed \textit{locally} behind the particle [see Fig.~1(a) of the main text], the equivalence between the global and local feedback protocols is not so obvious. In this section, we show the equivalence between these two feedback protocols from a topological point of view.

The feedback control considered in Ref.~\cite{Toyabe10} is as follows. A particle moves in a tilted periodic potential shown in the upper panel of Fig.~\ref{fig_Toyabe_sublattice}(a). The regions A and B (C and D) respectively denote the left and right sides of the local minimum (maximum) of the potential in the unit cell. If a particle is found in the region C, the phase of the periodic potential is shifted by $\pi$ so that the local minima and maxima are exchanged, thereby preventing the particle from sliding down the potential. To construct a lattice model for this setup, we discretize the periodic potential into four sites for each unit cell, which represent the regions A, B, C, and D [see the lower panel of Fig.~\ref{fig_Toyabe_sublattice}(a)]. The potential difference between sublattices A and B, the potential difference between neighboring unit cells, and the potential barrier between the local minimum and maximum are denoted by $\delta,\Delta_1$ and $\Delta_2$, respectively. We consider continuous time evolution of the probability distribution of a particle in this lattice model described by a master equation $\frac{d\bm{p}}{dt}=R\bm{p}$, where the transition rates are determined by the potential differences between neighboring sites so that local detailed balance conditions are satisfied, as in the main text. We perform a measurement of the position of this particle. If the particle is detected in sublattice A, B, or D, we do nothing and the particle undergoes the time evolution during time $\tau$ according to the master equation with the same rate matrix $R$. If the particle is detected in sublattice C, we switch the potential as in the lower right panel of Fig.~\ref{fig_Toyabe_sublattice}(b) and the particle undergoes the time evolution during time $\tau$ with the rate matrix modified accordingly.

In Fig.~\ref{fig_Toyabe_sublat_spec}, we show the eigenspectrum of the stochastic matrix of this model. Panel (a) [(b)] corresponds to the case without (with) feedback control. Reflecting the four-sublattice structure, the eigenspectra consists of four loops.
The winding numbers of the right-most loops in (a) and (b), which include the unit eigenvalues of steady states, have opposite signs around a reference point inside each loop. 
This behavior is topologically the same as that observed for the model shown in the main text [see Fig.~1(c) and (e)].

\section{Relationship between zero eigenvalue and Markovianity}
In this section, we elaborate on the relationship between Markovianity and zero eiganvalue of a stochastic matrix, thereby showing that the reference point at the origin of the complex plane plays a special role in the topological characterization of non-Markovianity. Suppose that a stochastic matrix $T$ is generated by a Markovian master equation as
\begin{equation}
T=\mathcal{T}\exp\left[\int_0^\tau dt R(t)\right],
\end{equation}
where $R(t)$ is a rate matrix and $\mathcal{T}$ is the time-ordering operator. Then, $T$ is invertible with its inverse given by $\tilde{\mathcal{T}}\exp\left[-\int_0^\tau dtR(t)\right]$, where $\tilde{\mathcal{T}}$ denotes the anti-time-ordering operator. Physically, invertibility of $T$ means that for any two distinct initial distributions $\bm{p}_1$ and $\bm{p}_2$ ($\bm{p}_1\neq\bm{p}_2$), the distributions after the time evolution are distinguishable: $T\bm{p}_1\neq T\bm{p}_2$. Conversely, if a stochastic matrix $T=(T_{i,a;j,b})$ has zero eigenvalue, there exist two probability distributions $\bm{p}_1$ and $\bm{p}_2$ such that $T\bm{p}_1= T\bm{p}_2$; in fact, if $T\bm{v}=0$ with $\bm{v}=(v_{j,b})$ being an eigenvector, we can write $\bm{v}\propto\bm{p}_1-\bm{p}_2$ since $\sum_{j,b} v_{j,b}=0$ follows from $\sum_{i,a}T_{i,a;j,b}=1$.

It is clear that non-Markovian processes can realize stochastic matrices with zero eigenvalue. For example, a stochastic matrix for a two-level system
\begin{equation}
T=
\begin{pmatrix}
    1 & 1 \\
    0 & 0
\end{pmatrix},
\end{equation}
which has zero eigenvalue, can be realized with feedback control by flipping the state (doing nothing) if we find the system at the lower (upper) level. This feedback control leads to the same final state regardless of the initial state, thereby rendering the distributions starting from different initial conditions indistinguishable. Note that feedback control is a non-Markovian process because the operation is conditioned on the pre-measurement state of the system. Thus, if a stochastic matrix has a nontrivial value of a topological invariant around the origin of the complex plane, it cannot be continuously deformed from the identity matrix without its eigenvalues crossing the origin and necessarily requires non-Markovianity. Feedback control produces such a desired effect.

Quantum effects enable realization of a stochastic matrix beyond the Markovian constraint. To see this, we consider a simple example of a stochastic matrix for a two-level system:
\begin{equation}
T=
\begin{pmatrix}
    0 & 1 \\
    1 & 0
\end{pmatrix}.
\end{equation}
This stochastic matrix has $s=\mathrm{sgn}(\det[T])=-1$ [see Eq.~\eqref{eq_detT}] and thus cannot be realized with any classical Markovian master equation because $s$ cannot be changed unless $T$ has a zero eigenvalue. However, it can be realized in the quantum case as \cite{Korzekwa21}
\begin{equation}
T=T(\pi/2),
\end{equation}
where
\begin{gather}
[T(\tau)]_{i,j}=\bra{i}\mathcal{E}_\tau(\ket{j}\bra{j})\ket{i},\label{eq_supple_T_qembed}\\
\mathcal{E}_\tau(\rho):=e^{-i\sigma_x \tau}\rho e^{i\sigma_x\tau},
\end{gather}
$j=0,1$ denotes the two levels, and $\sigma_x$ is the $x$ component of the Pauli matrix. In fact, we have
\begin{equation}
T(\tau)=
\begin{pmatrix}
    \cos^2\tau & \sin^2\tau \\
    \sin^2\tau & \cos^2\tau
\end{pmatrix},
\end{equation}
which has zero eigenvalue at $\tau=\pi/4$ as $\det T(\pi/4)=0$.

The distinction between classical and quantum dynamics in simulating stochastic processes can be understood in terms of zero eigenvalue. For quantum dynamics governed by a time-local quantum master equation $\frac{d}{dt}\rho(t)=\mathcal{L}(t)(\rho(t))$, the quantum channel $\mathcal{E}_\tau=\mathcal{T}\exp[\int_0^\tau dt\mathcal{L}(t)]$ does not have zero eigenvalue. Consequently, two solutions of the density matrix starting from different initial conditions are distinguishable: $\mathcal{E}_\tau(\rho_1)\neq\mathcal{E}_{\tau}(\rho_2)$ if $\rho_1\neq\rho_2$. However, the classical probability distributions $\bm{p}_\alpha=(P_j\rho_\alpha P_j)_j\ (\alpha=1,2)$ obtained from projection $P_j=\ket{j}\bra{j}$ onto the diagonal components of the density matrix may be indistinguishable. Therefore, a stochastic matrix with zero eigenvalue can be realized as Eq.~\eqref{eq_supple_T_qembed} with quantum Markovian dynamics. The emergence of zero eigenvalue can also be understood as a consequence of projective measurements since
\begin{align}
[T(\tau)]_{i,j}=&\bra{i}P_i\mathcal{E}_\tau(P_j\ket{j}\bra{j}P_j)P_i\ket{i}\notag\\
=&\bra{i}\mathcal{E}_{\mathrm{proj}}\circ\mathcal{E}_\tau\circ\mathcal{E}_{\mathrm{proj}}(\ket{j}\bra{j})\ket{i},
\end{align}
where $\mathcal{E}_{\mathrm{proj}}(\rho)=\sum_jP_j\rho P_j$ is the projective measurement channel, 
and the concatenated channel $\mathcal{E}_{\mathrm{proj}}\circ\mathcal{E}_\tau\circ\mathcal{E}_{\mathrm{proj}}$ has zero eigenvalue due to projection. Physically, the memory of the initial state is retained in the off-diagonal components of the density matrix, enabling the effective non-Markovian classical stochastic dynamics \cite{Korzekwa21}.

The topological transition in the classical stochastic matrix \eqref{eq_supple_T_qembed} is related to an exceptional point of the corresponding quantum channel.
To see this, we consider a one-parameter family of quantum channels
\begin{equation}
\mathcal{E}_\tau^\prime=\mathcal{E}_\tau\circ\mathcal{E}_{\mathrm{proj}}.
\label{eq_channel_homotopy}
\end{equation}
Since $\mathcal{E}_\tau$ is invertible, we have $\mathrm{dim}\ker[\mathcal{E}_\tau^\prime]=\mathrm{dim}\ker[\mathcal{E}_{\mathrm{proj}}]$. Thus, the degeneracy of zero eigenvalues is unchanged under the continuous deformation controlled by the parameter $\tau$.
The quantum channel $\mathcal{E}_\tau^\prime$ has four eigenvalues. Two of them are zero arising from off-diagonal components of the density matrix: $\mathcal{E}_\tau^\prime(\ket{0}\bra{1})=\mathcal{E}_\tau^\prime(\ket{1}\bra{0})=0$. The other two are the same as those of the stochastic matrix $T(\tau)$ defined in Eq.~\eqref{eq_supple_T_qembed}; in fact, by setting
\begin{equation}
\rho=
\begin{pmatrix}
\rho_{00} & \rho_{01} \\
\rho_{10} & \rho_{11}
\end{pmatrix},
\end{equation}
the eigenvalue problem $\mathcal{E}_\tau^\prime(\rho)=\xi\rho$ reduces to $T(\tau)\bm{p}=\xi\bm{p}$, where $\bm{p}=(\rho_{00},\rho_{11})^\top$ and the off-diagonal components are given as $\xi\rho_{01}=\xi\rho_{10}^*=i(\rho_{00}-\rho_{11})\cos\tau\sin\tau$. For $\tau=\pi/4$, the eigenvector of $T(\pi/4)$ with zero eigenvalue is given by $\rho_{00}=-\rho_{11}=1/2$, and the corresponding eigenoperator of the quantum channel is given by
\begin{equation}
\xi\rho=
\begin{pmatrix}
    0 & i/2 \\
    -i/2 & 0
\end{pmatrix},
\end{equation}
which is linearly dependent on the other eigenmodes with zero eigenvalues. Thus, $\mathcal{E}_{\pi/4}^\prime$ is not diagonalizable and located at an exceptional point \cite{Ashida20}. By forming the exceptional point, the eigenvalues of the quantum channel $\mathcal{E}_{\tau}^\prime$ can cross the origin of the complex plane as $\tau$ changes without altering the degeneracy of zero eigenvalues.

\end{document}